\documentclass[aps,prb,showpacs,twocolumn,floatfix]{revtex4}
\usepackage{amssymb}
\usepackage{graphicx}
\usepackage{dcolumn}
\usepackage{bm}
\usepackage[]{amsmath}

\begin{document}

\title{Orbital-spin ordering in the striped antiferromagnetic state of iron-based superconductors }
\author{Guo-Qiang Liu}
\affiliation{Max-Planck-Institut f\"ur Festk\"orperforschung, D-70569 Stuttgart, Germany}
\date{\today }

\begin{abstract}
The magnetic properties of iron-arsenides are investigated using
the LSDA+$U$ approach. In addition to one high moment state, we
find that a positive $U$ also produces two low moment states with
m$\thicksim$0.4 $\mu_{B}$ and m$\thicksim$1.0 $\mu_{B}$. The electronic structures
indicate that the low moment states originate in the strong orbital
hybridization between antiferromagnetic Fe neighbors. Therefore,
the geometry of the FeAs$_{4}$ tetrahedron is crucial to the low moment
states, which is the key to understand the negative pressure
coefficient of T$_{C}$ in LiFeAs. Finally, our theory suggests that the
superconducting phase is an orbital-spin ordered state, where the
orbital spin-moments cancel each other.
\end{abstract}
\maketitle

\section{Introduction}
Iron-based superconductors (FeSC) have attracted intense attention
since the discovery of superconductivity in LaFeAsO$_{1-x}$F$_{x}$
with T$_{C}$ of 26 K. \cite{Kamihara} A central topic of the
studies is to understand the mechanism behind superconductivity.
Since the electron-phonon coupling strength was proved to be
insufficient to explain the relative high T$_{C}$ both
theoretically and experimentally, \cite{Boeri,Boeri2,Christianson}
spin fluctuations are expected as an alternative candidate for the
pairing glue. Therefore understanding the magnetic behavior
of the parent compounds is urgently needed.

Unexpectedly, the parent compounds of FeSC exhibit a wide variety
of magnetic properties despite their similar crystal structures.
The undoped compounds are antiferromagnetic metals with the
moments ordered in stripes, i.e. ferromagnetically between the
nearest Fe neighbors in one direction (x) and
antiferromagnetically in the other (y). The moment of the 1111
compounds RFeAsO (R=rare earth) is $\thicksim$ 0.3-0.8 $\mu_{B}$.
\cite{Cruz,Zhao,Kimber,Qiu} In the 122 compounds
AFe$_{2}$As$_{2}$ (A=Ca, Sr, Ba, Eu) it is $\thicksim$ 0.8-1.0
$\mu_{B}$, \cite{Goldman,Zhao_J,Su} whereas in the 111 family the
moment is only $\thicksim$ 0.09 $\mu_{B}$ in NaFeAs \cite{Li} and
could not be observed in LiFeAs. \cite{Tapp} In the 11 family, no
local moment is found in FeSe, \cite{Medvedev} while FeTe has a
unique magnetic structure, a so called double-stripe order, with a
fairly big moment $\thicksim$ 2.0 $\mu_{B}$. \cite{Li_S} This
spread in magnetic moments is difficult to understand since the
parent compounds have very similar lattice structures; they all
have similar FeX (X=As, Se, Te) layers, and density functional
theory (DFT) calculations give quite similar paramagnetic band
structures for all the materials. \cite{Miyake}

It is conceivable that theoretical studies may clarify this
puzzle. DFT with the local spin density approximation (LSDA) or
the generalized gradient approximation (GGA) successfully
reproduced the single-stripe antiferromagnetic (AFM) order in 1111
and 122 materials. \cite{Ishibashi, Yin, Singh, Yaresko} However
the calculated magnetic moment is about 1.5-2 $\mu_{B}$, which is
much larger than the measured magnetic moment. Many theories have
been proposed to explain the outstanding discrepancy between DFT
and experiment. Yildirim applied a fixed-spin-moment calculation
to LaFeAsO and found the ground state with a moment of $m=0.48$
$\mu_{B}$. \cite{Yildirim} Nakamura $et$ $al$. applied the LSDA+U
approach to LaFeAsO$_{1-x}$F$_{x}$ and reproduced the experimental
moment with a $negative$ U. \cite{Nakamura} Mazin and Johannes
argued that the small moment is due to the presence of antiphase
boundaries. \cite{Mazin} Bascones $et$ $al$. presented a five-band
model and found a low moment state characterized by anti-parallel
orbital magnetic moments. \cite{Bascones} However, it is still not
clear which picture is more realistic.

In this work, we present an LSDA+$U$ study to explain the puzzle
of magnetism in FeSC. We show that the measured small magnetic
moment can be reproduced with $positive$ $U$, and its origin is
a Coulomb-enhanced orbital hybridization between
antiferromagnetic neighbors. Since the iron compounds are weak to
moderately correlated systems, \cite{Nakamura_K, Yang} the
around-mean-field (AMF) functional \cite{Czyzyk} is employed in
this work. All the calculations are
performed with the full-potential linear augmented plane wave
(FLAPW) within the local-density approximation, as implemented in
package WIEN2K. \cite{wien} We concentrate on the 1111 material LaFeAsO and
use the experimental lattice structure \cite{Kamihara} with the
space group P4/nmm, a=4.035 \AA, c=8.7409 \AA, z$_{La}$=0.1415 and
z$_{As}$=0.6415. To study the single-stripe AFM order, we use the
$\sqrt{2}\times\sqrt{2}\times1$ supercell containing 4 iron atoms.
144 k points are used within irreducible Brillouin zone. The
muffin-tin sphere radii of 2.24, 2.28, 2.02 and
1.98 a$_0$ are used for La, Fe, As and O, respectively.

\section{Results and discussion}
\subsection{Magnetism and band structure}
\begin{figure}[tbp]
{\scalebox{0.65}[0.65]{\includegraphics{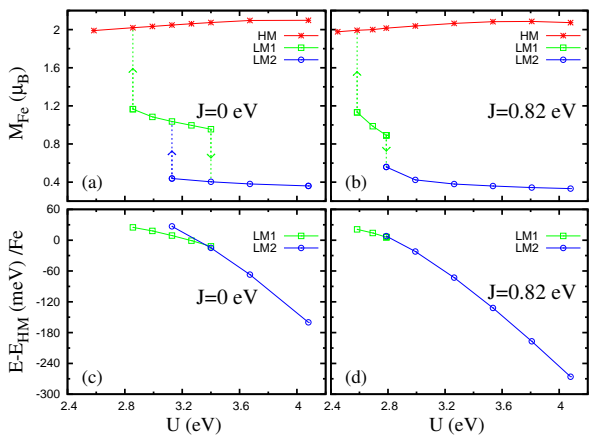}}}
\caption{Calculated magnetic moment and total energy for
LaFeAsO as a function of $U$.} \label{mom}
\end{figure}

Since LSDA+$U$ may give multiple minima, \cite{Erik} we change the
starting magnetic moment on Fe to search for all the solutions.
Fig. \ref{mom} shows our calculated local magnetic moment on Fe
and the total energy as a function of the Coulomb interaction $U$.
We used two values for the Hund's coupling, $J$=0 and $J$=0.82 eV,
the latter having been obtained from an X-ray experiment.
\cite{Yang} Surprisingly, two kinds of low moment (LM) in
addition to one high moment (HM) solutions are found in our
calculations. The HM state was expected since the LSDA+$U$ is
usually believed to have a tendency to enhance the magnetic
moment. The two LM states have the magnetic moments
m$\thicksim$1.0 $\mu_{B}$ (LM1) and m$\thicksim$0.4 $\mu_{B}$
(LM2), which are in good agreement with the measured moments in
the 122 compounds and in most of the 1111 compounds, respectively.
We did the same calculation for the 122 materials
BaFe$_{2}$As$_{2}$, and also found two LM states, very similar to
LaFeAsO.

As shown in Fig. 1a and 1b, the LM1 state only exists within a
certain $U$ range, and it can jump to the HM or LM2 states with a
discontinuous change of the magnetic moment when $U$ decreases or
increases. The LM2 state requires larger $U$ and shows slight
variation in moment when $U$ is increasing. Fig. 1c and 1d present
the total energy of the LM1 and LM2 states relative to the HM
state. The figure shows that a small $U$ favors the HM state and a
large $U$ favors the LM2 state. Consequently, the experimental
ground state of LaFeAsO \cite{Cruz} can be reproduced in our
calculation if $U$ is large enough. The effect of $J$ may also be seen
in Fig. 1. In general, the results for $J$=0 and $J$=0.82 eV are
similar. We therefore only consider the case of $J$=0 in the
following parts of this work.

\begin{figure}[tbp]
{\scalebox{0.65}[0.65]{\includegraphics{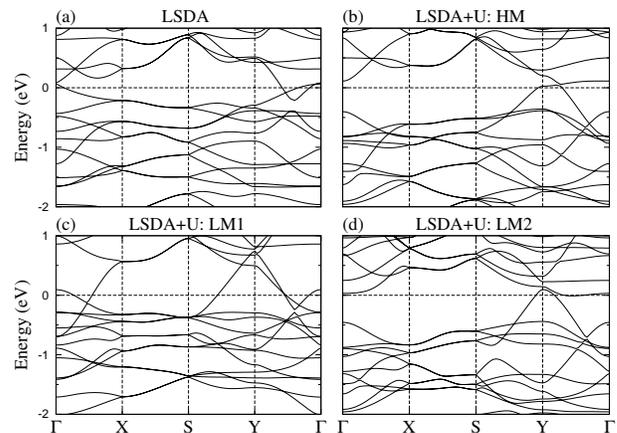}}} \caption{
Theoretical band structures of LaFeAsO for single-stripe AFM
state. All the LSDA+$U$ band structures are for $J$=0
and $U$=3.26 eV.} \label{bnds}
\end{figure}

Our theoretical band structures of LaFeAsO for the different
states are presented in Fig. \ref{bnds}. The LM1 and LM2 states
show very different band structures. The LM1 indicates a rather
good metal while the LM2 state is nearly semiconducting.
Therefore, the magnetic strength and electrical conductivity are
correlated properties in the iron compounds. The coexistence of
the LM1 and LM2 states in our calculations may clarify some
seemingly inconsistent experiments. McGuire $et$ $al$.
\cite{McGuire} reported a systematic study on the 1111 materials.
They found that CeFeAsO has a magnetic moment about 0.3 $\mu_{B}$ and
its low temperatures electrical resistivity is $\rho_{0}$=2
m$\Omega$ cm. While neutron scatting experiment by Zhao $et$ $al$.
\cite{Zhao} indicated a magnetic moment of 0.8 $\mu_{B}$ for
CeFeAsO, and Jesche $et$ $al$. \cite{Jesche} reported an
electrical resistivity of $\rho_{0}$=140 $\mu\Omega$ cm for
CeFeAsO. It should be noticed that McGuire $et$ $al$. used
polycrystalline samples and Jesche $et$ $al$. used large single
crystals. Moreover, a recent neutron scatting experiment
\cite{Li_HF} found single crystal LaFeAsO has a moment of 0.8
$\mu_{B}$, which is also very different from the earlier study.
\cite{Cruz} From our theory, these discrepancies can be naturally
explained by assuming that the samples were in different states,
which may be due to the quality of samples.

\subsection{Origin of the low moment states}
In the LSDA+$U$ calculations, the on-site elements of the density
matrix may contain some important information about the detail of
the states. In Table \ref{occ}, we present the occupancy of the Fe
$3d$ orbitals and the contribution of each orbital to the spin
magnetic moment. The main elements of the spin-dependent local
density matrices can be extracted from this table. The main
difference between the three LSDA+U states lies in the moment of
the $xy$ orbital. In the HM state, $m_{xy}$ is enhanced from the
LSDA value of 0.42 $\mu_{B}$ to 0.52 $\mu_{B}$. But in the LM1 and
LM2 states, $m_{xy}$ is reduced to 0 and -0.43 $\mu_{B}$! This
unusual reduction implies that the $xy$ orbital is the key to
understand the LM states. The HM, LM1 and LM2 states may be
characterized as $m_{xy}$$>$0, $m_{xy}$$\thicksim$0 and
$m_{xy}$$<$0, respectively. We may therefore name the LM states as
'orbital-spin ordered' states. In the LM2 state, $m_{xy}$ and
$m_{yz}$ have opposite signs from $m_{x^{2}-y^{2}}$ and $m_{xz}$.
It is worth noting that this result is in very good agreement with
the tight-binding study \cite{Bascones} by Bascones $et$ $al$.

\begin{table}[!htbp]
\centering \caption{ Occupancy of the Fe 3d orbitals and
orbital contribution to spin magnetic moment. The data
for different solutions correspond to the band structures
in Fig. 2. All the LSDA+$U$ data are obtained with $J$=0 and $U$=3.26 eV.}
\begin{tabular*}{8.2cm}{c@{\extracolsep\fill}c@{\extracolsep\fill}c@{\extracolsep\fill}c@{\extracolsep\fill}c@{\extracolsep\fill}c@{\extracolsep\fill}c}
\hline\hline
   & $3z^{2}-r^{2}$ & $x^{2}-y^{2}$ & xy & xz& yz& total\\
 \hline
 \multicolumn{2}{c} {LSDA}\\
\hline
occupancy   &1.29    & 1.19    & 1.21  &   1.15   &  1.34 &6.19\\
 moment       &0.35    & 0.30   &  0.42  &   0.44 &    0.25& 1.76\\
\hline
  \multicolumn{2}{c} {LSDA+$U$: HM}\\
 \hline
 occupancy   & 1.25 &     1.16  &     1.19  &    1.09 &     1.47& 6.16\\
 moment       & 0.43&     0.42 &     0.52 &    0.56&     0.13 & 2.05\\
\hline
 \multicolumn{2}{c} {LSDA+$U$: LM1}\\
\hline
occupancy  &1.37 &     0.98  &   1.45 &    1.06  &   1.33 &  6.19\\
moment       &0.06&     0.28   &  0.00  &   0.57   &  0.08 & 0.99 \\
 \hline
 \multicolumn{2}{c} {LSDA+$U$: LM2}\\
 \hline
 occupancy  & 1.36 &     1.05  &   1.24 &    1.08  &   1.44& 6.18\\
 moment       & 0.03 &     0.31  &  -0.43  &   0.55  &  -0.05 & 0.41\\
\hline\hline
\label{occ}
\end{tabular*}
\end{table}
Compared with the LSDA solution, LSDA+$U$ rearranges the orbital
occupancy and orbital spin-moment. The LSDA+$U$ method can be
mapped onto a single-band picture if there is no strong orbital
hybridization. In this case, the Coulomb interaction has two
possible effects on the magnetic moment. The first is to enhance
the exchange splitting between the majority spin and minority
spin, which causes larger moment for each orbital. The second is
to induce a charge transfer from the less occupied orbitals to
more occupied orbitals, which may decrease the moment of the most
occupied orbital. As may be seen, the HM solution agrees well with
this picture, but the strong rearrangement of $m_{xy}$ and
$m_{3z^{2}-r^{2}}$ in the LM solutions cannot be understood. The
deficiency of the single-band picture suggests that interorbital
coupling plays an important role in the LM states. This has been
realized by Bascones $et$ $al$. \cite{Bascones} They stressed that
it is important to include all five $d$ orbitals in model studies.

\begin{figure}[tbp]
{\scalebox{0.65}[0.65]{\includegraphics{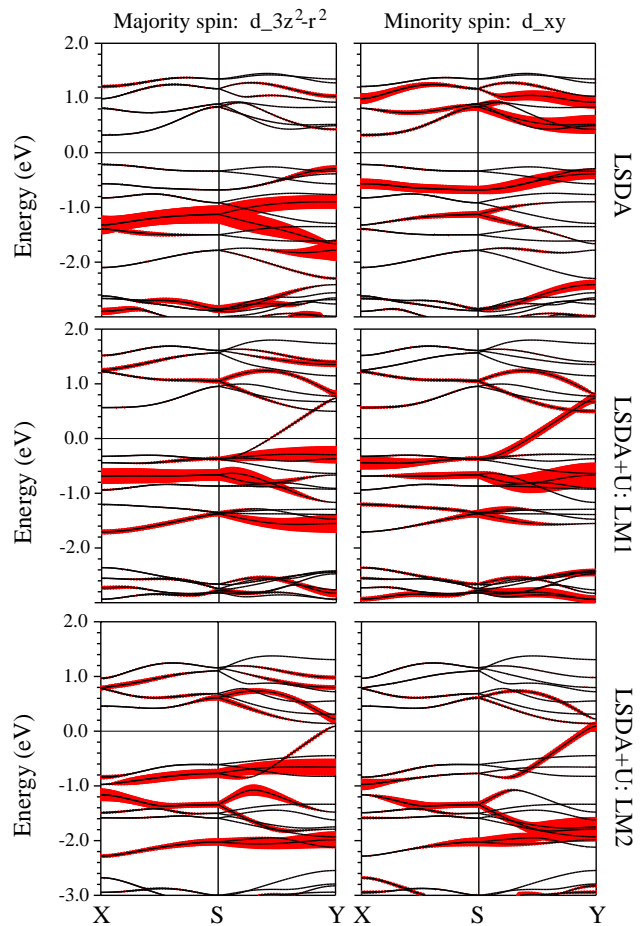}}} \caption{
Band character for the LSDA, LM1 and LM2 solutions. The
left panels are for the $3z^{2}-r^{2}$ orbital with majority spin,
the right panels for the $xy$ orbital with minority spin.} \label{fatband}
\end{figure}

Fig. \ref{fatband} shows the band character for the
$d_{3z^{2}-r^{2}}^{\sigma}$ and $d_{xy}^{\bar{\sigma}}$ orbitals,
where $\sigma$ stands for the majority spin and $\bar{\sigma}$ for
the minority spin. One may see that the hybridization of these two
orbitals is strongly enhanced in the LM states relative to the
LSDA solution. This strong hybridization between different spin
channels (or more precisely, between antiferromagnetic neighbors)
rationalizes the reduction of $m_{xy}$ and $m_{3z^{2}-r^{2}}$ in
the LM states. Similar enhanced hybridization is also found
between some other antiferromagnetic orbitals, such as
$d_{x^{2}-y^{2}}^{\sigma}$-$d_{xy}^{\bar{\sigma}}$ and
$d_{yz}^{\sigma}$-$d_{xy}^{\bar{\sigma}}$. Here we focus on this
$d_{3z^{2}-r^{2}}^{\sigma}$-$d_{xy}^{\bar{\sigma}}$ hybridization.
The strength of orbital hybridization depends on two factors:
hopping integral and orbital on-site energy. In the LSDA+$U$
approach, $U$ does not change the hopping integrals, but shifts
the on-site energies. As shown in Fig. \ref{fatband}, the
$d_{3z^{2}-r^{2}}^{\sigma}$ and $d_{xy}^{\bar{\sigma}}$ bands are
much closer in the LM solutions than those in the LSDA solution,
which is responsible for the enhancement of orbital hybridization.

In a tight-binding
model, the $d$ orbital on-site energy can be
written as
\begin{equation}
\label{onsite}
\epsilon_{\alpha}^{\sigma}=\epsilon_{\alpha}^{0}\pm\frac{1}{2}Im-Un_{\alpha}^{\sigma}+\epsilon_{dc}^{\sigma}.
\end{equation}
where $\epsilon$ stands for the on-site energy, $\alpha$ for the
orbital, $\sigma$ for the spin, $m$ for the local magnetic moment,
$n$ for the occupancy number. $I$ is the Stoner exchange integral,
which is about 0.9 eV in metallic Fe. \cite{And} The minus (plus)
sign of the term $\frac{1}{2}Im$ is for the spin that is parallel
(anti-parallel) to the direction of the local magnetic moment.
$\epsilon_{dc}^{\sigma}$ corresponds to the double counting term
of the LSDA+$U$ method. In this work, we use the AMF functional \cite{Czyzyk}
and have
$\epsilon_{dc}^{\sigma}=U\sum_{\alpha}n_{\alpha}^{\sigma}/5$ for
the $d$ orbitals. For two orbitals with opposite spins, we get
\begin{equation}
\label{diff}
(\epsilon_{\alpha}^{\sigma}-\epsilon_{\beta}^{\bar{\sigma}})=-Im-U(n_{\alpha}^{\sigma}-n_{\beta}^{\bar{\sigma}})+(\epsilon_{dc}^{\sigma}-\epsilon_{dc}^{\bar{\sigma}})+(\epsilon_{\alpha}^{0}-\epsilon_{\beta}^{0}).
\end{equation}
It is reasonable to assume
$(\epsilon_{3z^{2}-r^{2}}^{0}-\epsilon_{xy}^{0})=0$, as the LSDA
solution shows that the $3z^{2}-r^{2}$ and $xy$ orbitals have
similar occupancies. Then we can calculate the gap between the
$d_{3z^{2}-r^{2}}^{\sigma}$ and $d_{xy}^{\bar{\sigma}}$ orbitals.
Using the data in Table \ref{occ}, we get -1.58 eV for the LSDA,
-0.22 eV for the LM1 and 0.35 eV for the LM2. These values are
consistent with the character bands shown in Fig. \ref{fatband}. To
enhance the hybridization, the $d_{3z^{2}-r^{2}}^{\sigma}$ and
$d_{xy}^{\bar{\sigma}}$ orbitals need to be close. Eq. \ref{diff}
shows some competitive factors for this requirement. The term
$-Im$ opens a gap between opposite spins, and therefore the $U$
term and the double counting term need to offset this splitting.
This gives a restriction on $U$. As shown in Fig. \ref{mom}, there
is no LM solution when $U$ is too small. For the LSDA state, the
exchange splitting $-Im$ opens a gap of -1.58 eV. If we start the
LSDA+$U$ calculation with the LSDA solution, this gap is too large
to be overcomed and the system is easier to converge to the HM state.
To search for the LM solutions, we need to start the LSDA+$U$
calculation with a small magnetic moment. This may be the
technical reason why the LM states were not found before.

\begin{figure}[tbp]
{\scalebox{0.65}[0.6]{\includegraphics{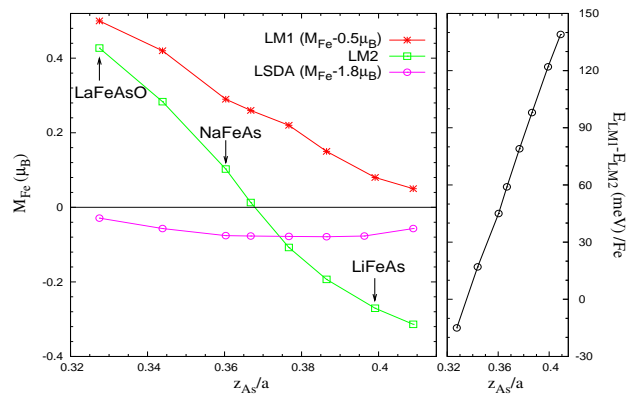}}} \caption{
Magnetic moment and total energy of LaFeAsO as a function of $c/a$
with constant volume. The magnetic moment for LM1 and LSDA are
shifted to fit the window. The experimental values of $z_{As}/a$
for LaFeAsO, NaFeAs and LiFeAs are indicated by arrows. The
LSDA+$U$ data are calculated with $J$=0 and $U$=3.26 eV. }
\label{coa}
\end{figure}

We have shown that the role of $U$ in the LM states is to enhance
the orbital hybridization between antiferromagnetic neighbors by
changing the on-site energies. Here we discuss the effect of
$d$-$d$ interorbital hopping, which is the other factor
influencing the strength of hybridization. From the structure of
LaFeAsO, we may realize that the direct interaction between the
$d_{xy}$ and $d_{3z^{2}-r^{2}}$ orbitals is weak, and therefore
the hybridization of $d_{3z^{2}-r^{2}}^{\sigma}$ and
$d_{xy}^{\bar{\sigma}}$ is via the As $p$ orbitals. These $p$-$d$
interactions, such as $p_{z}$-$d_{xy}$ and
$p_{z}$-$d_{3z^{2}-r^{2}}$, are very sensitive to the Fe-As-Fe
angle. This characteristic can be used to tune the $d$-$d$
interorbital hoppings. We adjust the Fe-As-Fe angle by changing
the ratio between lattice constant $c$ and $a$ while keeping the
volume constant. Our calculated magnetic moment of LaFeAsO as a
function of $c$/$a$ is shown in Fig. \ref{coa}. The $x$-axis of
the figure is labelled $z_{As}/a$, where $z_{As}$ is the height of
the As atom above the nearest Fe layer. In the LM2 state, $m_{xy}$
is anti-parallel to the direction of the total moment at small
$z_{As}/a$ as shown in Table \ref{occ}. As $z_{As}/a$ increases,
$m_{xy}$ becomes numerically larger and eventually the total
moment gets the same direction as $m_{xy}$, which is indicated by
the negative sign in Fig. \ref{coa}. The figure shows that the
LSDA state has a very weak dependence on the value of $z_{As}/a$.
Please note that the Fe-As distance almost keeps a constant in
this calculation. On the other hand, the LM states show very
strong dependence on the FeAs$_{4}$ geometry. This phenomenon
confirms the importance of the interorbital coupling to the LM
states. We may conclude that the origin of the LM solutions is the
Coulomb-enhanced orbital hybridization between antiferromagnetic
neighbors.
\subsection{Low moment solution from GGA}
Fig. 1 indicates a critical value $U_{c}$ for the LSDA+$U$
calculations: the LM states only exist at $U>U_{c}$. According to
our explanation for the LM states, the Coulomb interaction and
FeAs$_{4}$ structure are the two factors giving rise to the
orbital-spin ordered states. Then $U_{c}$ may be reduced by
changing the FeAs$_{4}$ geometry. We find that $U_{c}$ for NaFeAs
is about 1.9 eV, which is much smaller than the value for LaFeAsO
($\thicksim$2.8 eV). An interesting question is raised: can LSDA
or GGA produce the LM states without $U$?

Fig. 5 shows the low moment solution we find with GGA (without
$U$). The GGA-LM state appears at $z_{As}/a$=0.40, and the total
magnetic moment reaches a maximum at $z_{As}/a$=0.46. The GGA-LM
solution is not a simple low moment state, but an orbital-spin
ordered state. At $z_{As}/a$=0.46, the orbital moments are
$m_{3z^{2}-r^{2}}$=0.02, $m_{x^{2}-y^{2}}$=0.04, $m_{xy}$=-0.23,
$m_{xz}$=0.33 and  $m_{yz}$=0.01 $\mu_{B}$. We also present the
band structure for $z_{As}/a$=0.46 in Fig. 5. The very different
band dispersion along $x$ direction and $y$ direction shows that
this state is far away from the paramagnetic phase although it has
a small total moment. As may be seen, the GGA-LM band structure is
a little bit similar to the LM2 band structure shown in Fig. 2,
while the main difference is the Coulomb interaction nearly opens
a gap at Fermi level. Interestingly, we find that LSDA can hardly
produce this low moment state. A possible reason is that LSDA
contains less electron correlation in the iron compounds than GGA
does.

The LSDA+$U$ method is sometimes questioned due to the double
counting problem. \cite{Erik, Petukhov} The choice of LSDA+$U$
functional is somewhat arbitrary. The GGA-LM state proves that the
orbital-spin ordering does not come from a specific LSDA+$U$
functional, but originates in the tetrahedral structure of
FeAs$_{4}$.  The GGA-LM state provides a strong evidence for the
orbital-spin ordering in FeSC and also confirms our explanation
for the LM states.
\begin{figure}[tbp]
{\scalebox{0.67}[0.61]{\includegraphics{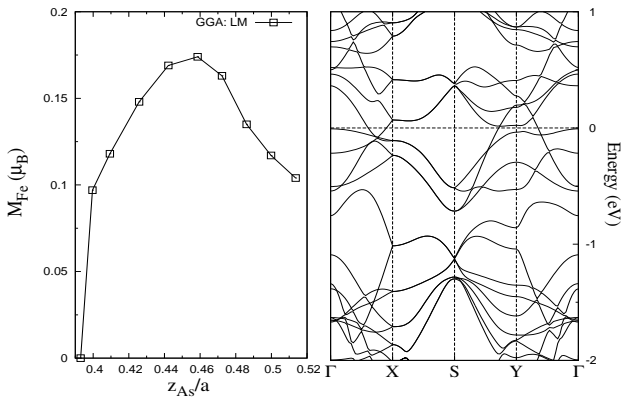}}}
\caption{Low moment state in LaFeAsO calculated by GGA (WITHOUT
$U$). The left panel is the magnetic moment as a function of $c/a$
with constant volume. The right panel is the band structure for
$z_{As}/a=0.46$, which is corresponding to an orbital-spin ordered
state with $m_{tot}=0.17$, $m_{xy}=-0.23$ and $m_{xz}=0.33$
$\mu_{B}$. } \label{gga}
\end{figure}
\subsection{FeAs$_{4}$ geometry, magnetism and high-pressure experiments}
It is known that the FeAs$_{4}$ tetrahedron tends to have similar
volume in different iron-arsenides, while its geometry is tuned by
different inserted layers (see Table I of Ref. 16). For example,
LiFeAs has the largest As height $z_{As}$ in all iron-arsenides
while its lattice constant $a$ is the smallest. Therefore, Fig.
\ref{coa} allows us to estimate the magnetic moments of the 111
materials. The figure gives a moment of 0.1 $\mu_{B}$ for NaFeAs
and -0.27 $\mu_{B}$ for LiFeAs. Here we consider the LM2 state
because the LM2 state becomes more and more energetically
favorable when $z_{As}/a$ is increasing as shown in Fig.
\ref{coa}. To check these estimated values, we did a direct
calculation using the experimental structures, which gives
$m$=0.04 $\mu_{B}$ for NaFeAs and $m$=-0.20 $\mu_{B}$ for LiFeAs.
This agreement between the estimation and calculation shows that
the FeAs$_{4}$ geometry is a very important factor controlling the
magnetic properties of FeSC, which is responsible for the variety
of magnetic moments in the 1111 and 111 materials. Our calculated
magnetic moment for NaFeAs, $m$=0.04 $\mu_{B}$, is in good
agreement with the measured value. \cite{Li} But in contrast to
the experiment, \cite{Tapp} the ground state of LiFeAs in our
calculation has the single-stripe AFM order. Actually, there might
be experimental evidence for the striped AFM order. Borisenko $et$
$al$. measured the Fermi surface of LiFeAs using angle-resolved
photoemission. \cite{Borisenko} They found that the pocket around
the $\Gamma$ point is shaped like a butterfly, which indicates
that the 4-fold rotation symmetry is broken. This symmetry
breaking might be due to the striped AFM order.

Fig. 4 may also explain the behavior of the parent compounds under
pressure. High-pressure experiment shows that the undoped LaFeAsO
becomes superconducting at pressure $\thicksim2$ GPa and reaches a
maximum T$_{C}$ of 21 K at $\thicksim12$ GPa. For higher pressures
the T$_{C}$ decreases slowly. \cite{Okada} LiFeAs shows very
different pressure behavior from LaFeAsO. LiFeAs is a
superconductor without external pressure. Under pressure, the
T$_{C}$ of LiFeAs decreases linearly at a rate of 1.5 K/GPa,
\cite{Gooch} which means that the maximal T$_{C}$ is at zero
pressure. Interestingly, a different high-pressure experiment
indicated that the behavior of NaFeAs is more similar to LaFeAsO
than LiFeAs, \cite{Zhang} even though NaFeAs and LiFeAs belong to
the same family.

To explain these experiments, we need to consider the structural
distortion under pressure. Pressure experiments have shown that
the Fe-As bond length is very rigid while the Fe-Fe distance
shrinks much faster when pressure increases, \cite{Kimber_S, Mito}
i.e., $z_{As}/a$ increased by pressure. It is found that chemical
doping and pressure can both induce superconductivity in
BaFe$_{2}$As$_{2}$ and they have very similar influence on the
FeAs$_{4}$ geometry. \cite{Kimber_S, Rotter} A noticeable fact is
that chemical doping hardly changes the crystal volume.
\cite{Rotter} This finding shows that the FeAs$_{4}$ geometry is
the key structural feature to superconductivity,\cite{Kimber_S}
whereas the volume reduction under pressure is less important.
Therefore the $x$-axis direction of Fig. \ref{coa} can be regarded
as the direction of increasing pressure.

We consider the LM2
state in Fig. 4. For LaFeAsO and NaFeAs, pressure first suppresses the
magnetism. After the magnetic moment reaches 0, the pressure
starts to enhance the magnetism. It is known that the emergence of
superconductivity in the FeSC is related to the suppression of
magnetism. Then the decreasing magnetism corresponds to the
increasing T$_{C}$, conversely, the increasing magnetism
corresponds to the decreasing T$_{C}$. And the maximal T$_{C}$
occurs at the magnetic moment $m$=0. For LiFeAs, pressure can only
enhance the magnetism as shown in Fig. \ref{coa}. Consequently,
the T$_{C}$ of LiFeAs has a negative pressure coefficient,
$dT_{C}/dP<0$. Therefore, these different high-pressure
experiments can be consistently explained by our theory. Fig.
\ref{coa} shows that the LM2 state is more and more energetically
favorable than the LM1 state when $z_{As}/a$ increases; hence the
122 compounds, which have the LM1 ground state at 0 pressure, will
enter into the LM2 state when pressure increases. Then our
explanation for the high-pressure experiments is also suitable for
the 122 materials. Importantly, in this picture, the suppression
of magnetism is because the orbital spin-moments cancel each
other. This suggests the superconducting phase is not a 'real'
nonmagnetic state but an orbital-spin ordered state.
\subsection{LSDA+$U$ functionals}
In this work, we used the AMF LSDA+$U$ functional. We also checked
our calculation with another widely used scheme: the fully
localized limit (FLL) functional, \cite{Anisimov} which is
suggested to be more suitable for strongly correlated systems. We
found one LM solution coexists with a HM solution in this
functional. The LM solution is very similar to the LM2 state and
shows very similar $z_{As}/a$ dependence. However, we didn't find
the LM1 solution, and FLL energetically favors the HM solution.
\section{Summary}
In summary, we have studied the magnetic properties of
iron-arsenides using the LSDA+$U$ approach. Two low moment states
are found in our calculation, which reproduce the measured
magnetic moments in the 1111 and 122 materials well. The LM2 state
 shows good agreement with the low moment state found by previous
tight-binding study. Our study reveals that the origin of the low
magnetization is the strong orbital hybridization between
antiferromagnetic neighbors. From the picture indicated above we
have learned: Firstly, the geometry of FeAs$_{4}$ is crucial to
superconductivity and magnetism because it controls the $d$-$d$
interorbital coupling. Secondly, our theory suggests that the
superconducting phase is not a conventional nonmagnetic state, but
a state where the orbital spin-moments cancel each other. This
implies an inherent relationship between the magnetic correlation
and superconductivity.

\begin{acknowledgements}
The author gratefully acknowledges O. K. Andersen, O. Jepsen, L.
Boeri and A. N. Yaresko for helpful discussions and useful
comments.
\end{acknowledgements}

\end{document}